\documentclass[sigconf,nonacm]{acmart}
\AtBeginDocument{%
  \providecommand\BibTeX{{%
    \normalfont B\kern-0.5em{\scshape i\kern-0.25em b}\kern-0.8em\TeX}}}

\begin{document}

\title[Stance Inference in Twitter through GCN Collaborative Filtering with Minimal Supervision]{Stance Inference in Twitter through Graph Convolutional Collaborative Filtering Networks with Minimal Supervision}

\author{Zhiwei Zhou}
\affiliation{%
  \department{L3S Research Center}
  \institution{Leibniz Universität Hannover}
  \country{Germany}
}
\email{zzhou@l3s.uni-hannover.de}
\orcid{0000-0002-6592-1113}

\author{Erick Elejalde}
\affiliation{%
  \department{L3S Research Center}
  \institution{Leibniz Universität Hannover}
  \country{Germany}
}
\email{elejalde@l3s.uni-hannover.de}
\orcid{0000-0002-4755-1606}


\begin{abstract}
  Social Media (SM) has become a stage for people to share thoughts, emotions, opinions, and almost every other aspect of their daily lives. This abundance of human interaction makes SM particularly attractive for social sensing. Especially during polarizing events such as political elections or referendums, users post information and encourage others to support their side, using symbols such as hashtags to represent their attitudes. However, many users choose not to attach hashtags to their messages, use a different language, or show their position only indirectly. Thus, automatically identifying their opinions becomes a more challenging task. To uncover these implicit perspectives, we propose a collaborative filtering model based on Graph Convolutional Networks that exploits the textual content in messages and the rich connections between users and topics. Moreover, our approach only requires a small annotation effort compared to state-of-the-art solutions. Nevertheless, the proposed model achieves competitive performance in predicting individuals' stances. We analyze users' attitudes ahead of two constitutional referendums in Chile in 2020 and 2022. Using two large Twitter datasets, our model achieves improvements of 3.4\% in recall and 3.6\% in accuracy over the baselines.
\end{abstract}





\maketitle

\section{Introduction}
Digital social networks have been targeted as valuable data sources for social studies. 
Twitter, in particular, is one of the preferred options by researchers and practitioners due to its popularity (353M active users in 2023\footnote{\url{https://www.statista.com/statistics/303681/twitter-users-worldwide/}}) and flexible application programming interface (API). 
Researchers use Twitter as a tool to analyze social phenomena. Studies range from examining mass media attention and stock market movements to predicting political elections~\cite{elejalde2018nature,paul2017compass,di2021content}.
A fundamental problem in analyzing social media and how it might influence real-life events is identifying users' stances toward a topic of interest. For example, accurately predicting the perspective of large communities could help us understand political and social movements, poll elections, or improve marketing strategies.

Previous studies on understanding users' perspectives usually start by filtering content to the targeted topic~\cite{hoang2013politics} by using keyword-based or other rule-based approach~\cite{graells2020every}. However, in filtering users based on the usage of keywords, digital social media studies risk incurring selection bias. Moreover, these methods usually involve an expensive annotation process or rely on a sentiment polarity analysis (which often does not equal stance \cite{Wildemann2023}). Furthermore, they disregard valuable user information from discussions on other topics and other social information (e.g., tags, friends, endorsements, or profiles). For this last issue, research in recommendation systems offers a practical alternative \cite{wang2019silent, he2017neural}.
For example, a user might never express her political preferences on the platforms for a specific election. Instead, she limits herself to reading and maybe retweeting some posts from her favorite politician during the election period. Therefore, we need to capture her preferences by looking into other actions (e.g., retweets or connections) \cite{tan2011user,mcpherson2001birds}. Previous studies have shown that by looking into social interactions, it is also possible to capture the lean of users on a topic~\cite{xiao2020timme}. So, in this study, we investigate how social media features such as social connection and topic interactions can improve a content-based collaborative filtering approach for users' stance prediction. 

We proposed a graph convolutional network (GCN) based model that leverages several types of network relations to predict users' attitudes. Also, since the model is topic agnostic, it requires limited human annotation and is only for the final analysis stage. Moreover, social networks offer a rich space for discussions about diverse and often controversial topics. Thus, the users' opinions can go in many directions and represent multiple perspectives~\cite{10.1145/3201064.3201076}. Our proposed methodology represents users' positions in an embedding space that allows us to discover their affinity to different viewpoints rather than just a positive/negative stance on a topic. Our model achieves state-of-the-art performances, showing improvements of 3.4\% in recall and 3.6\% in accuracy over the baselines.

Our main contributions can be summarized as follows:
\begin{itemize}
    \item We introduce a semi-supervised deep learning model that allows inferring Twitter users' stances on multiple topics simultaneously. The minimal annotation required for the topics makes it easier to integrate into a practical application.
    \item We propose an encoding method for users positioning concerning the identified topics. This embedded space allows the exploration of opinions at different granularity levels (e.g., stance, user, community). 
    \item We conduct extensive experiments on large Twitter datasets and show that our method achieves competitive results. 
\end{itemize}

\section{State of the art}
In this Section, we review prior research on opinion mining, user similarity, and collaborative filtering, as these techniques relate to the challenges we face.

\subsection{Stance detection}
Stance detection is the task of automatically identifying the opinions of an individual or community on a specific topic. It offers a consistent approach to analyzing large volumes of unstructured data. These algorithms have been used to study from congressional session transcripts \cite{burfoot2011collective} to online forums debates \cite{anand2011cats,sridhar2014collective}. In the last decade, digital social media have attracted the attention of researchers studying people's opinions as most of the public debate has moved and concentrated on these platforms. In particular, Twitter is an appealing data source due to its large user base and active discussions on various topics. However, with short documents, informal language, and slang, social media content poses new challenges for traditional models of opinion mining~\cite{baldwin-etal-2013-noisy}. In response, recent studies have focused on other characteristic elements of social media. For example, people may express their stances on abortion through colored variations of heart emojis \cite{graells2020every}. Similarly, communities on opposite sides of a discussion usually adopt hashtags that represent their stance (e.g., \#TrudeauMustGo or \#Trudeau4MoreYears, \#ISISisNotIslam or \#DeportAllMuslims) \cite{jackson2015hijacking, xu2020hashtag}. Previous research has often used hashtags to extracting users' stance \cite{10.1145/2908131.2908150, kobellarz2022reaching, vilella2020immigration, di2021content}. In our analysis, we also exploit these distinguishing hashtags from different camps to profile Twitter users by simultaneously learning embeddings for users and hashtags.

\subsection{User homophily}
Given the dynamism and diversity that characterizes discussion on social media, relying only on users' use of certain hashtags, keywords, or other strict filtering rules may limit our observations and bias our results. We need to appeal to other features that will allow us to accurately infer their opinion in a given matter. Previous work has suggested that the principle of homophily~\cite{mcpherson2001birds}, where we assume that social entities will associate with similar others, can help us in this task. In Twitter, researchers have used networks based on the `following' relationship (both unidirectional and bidirectional) \cite{tan2011user} as well as the second-order co-following~\cite{garimella2014co}. 

Other Twitter features can also be used to create similarity networks. For example, in \cite{volkova2014inferring}, the authors propose (i) {\em Social Graph} - including different social circles such as friend or mention; (ii) {\em Entity-Centric Graph}, based on co-following relations between the users around a particular type of entity such as political candidates; and (iii) {\em Geo-Centric Graph}, grouping users with a given geopolitical profile, e.g., as self-reported in their biographies~\cite{volkova2014inferring}. Our analysis includes multiple relations between users from these three groups.

Furthermore, we leverage practical content elements that can also be used to establish relationships between users (e.g., hashtags, replies, retweets, and mentions (@))~\cite{tan2011user, volkova2014inferring, yang2013community}, even if these links are beyond the scope of our topic $i$ of interest.
Another relevant set of features is based on latent relations between users (i.e., not directly observable). For example, in \cite{sun2011pathsim}, the authors propose a sequence of relations defined between different object types to create a meta-path connecting similar objects (called {\em PathSim}).
With PathSim, we can identify objects that are strongly connected or share similar visibility in the network. 

Most current approaches aim at a stance polarity classification (i.e., in favor - against). Moreover, they rely on supervised learning, which makes these strategies difficult/costly to scale and deploy in practical scenarios. Our methodology contributes to this line of research by predicting the users' stance in a continuous higher-dimensional space, thus allowing a finer-grain stance analysis (i.e., not limited to polarity). Also, the minimal manual processing required offers a pipeline that is easier to apply in practical systems.
\subsection{Collaborative Filtering}
Most of the previous work on opinion mining focus on training a classifier. However, this task can also be framed from the recommendation system~(RS) perspective. In this case, we are interested in predicting user-topic affinity, or more precisely, user-[opinion on a topic]. There are two popular approaches in the area RS, namely: Matrix Factorization based methods and user-item graph structures analysis. Matrix factorization~(MF) projects the ID of a user $u$ (or an item $f$) into an embedding vector $H_u$ (or $W_f$)~\cite{koren2009matrix}. The missing user-item interactions are estimated by the inner product of $H_u$ and $W_f$. Some frameworks have tried to extend MF, e.g., by combining it with a multilayer perceptron (named neural collaborative filtering - NCF) \cite{he2017neural}. However, with the proper setting, the original MF method outperformed the NCF framework and other methods in most cases \cite{rendle2020neural, anelli2021reenvisioning}. Also, \citeauthor{wang2019silent} proposed a coupled sparse matrix factorization (CSMF) approach to collaborative filtering in the prediction of sentiments towards topics \cite{wang2019silent}. The authors relied on manually selected and annotated topics and used the accuracy of the sentiment polarity predictions to evaluate the model.

Alternatively, RS can be approached by exploiting the user-item bipartite graph structure. This creates a mapping from the RS to the link prediction problem. Motivated by the strength of graph convolution, \citeauthor{wang2019neural} proposed a Neural Graph Collaborative Filtering (NGCF) framework that captured collaborative signals in high-hop neighbors and integrates them into the embedding learning process \cite{wang2019neural}. However, further studies showed that NGCF demonstrates higher training loss and worse generalization performance with nonlinear activation and feature transformation \cite{he2020lightgcn}. As a result, the authors proposed a simplified model named Light Graph Convolution Network (LightGCN). Other works have leveraged LightGCN by aggregating information from different aspect-level graphs \cite{mei2021light} (e.g., adding a user-director graph on a user-movie recommendation to guide the embedding learning process). These RS models typically aggregate information by averaging data from neighbors. Alternatively, attention mechanisms have been also proposed to capture the importance of different relationships between users and items~\cite{fan2019graph}. In our experiments, we also test various aggregation strategies (see Section \ref{sec:effect_inferred_info})

\section{Datasets}
\label{sec:dataset}
This section describes the datasets used to train and validate our models. We start by presenting the case study and contextualizing the collected data. Then we define the collection process and the filters applied to the data, resulting in our final corpus. 

Following the FAIR data principles, we make our datasets available on GitHub\footnote{\url{https://github.com/imzzhou/StanceInferenceInTwitter.git}}. To comply with Twitter's terms and conditions, we only share tweet IDs that can be rehydrated.

\subsection{Case-study: Chilean constitutional referendum}
\label{chile_referendum}
In 2019, Chile saw one of its biggest popular uprisings following a perceived increase in economic hardship and social inequalities. After weeks of protest, lawmakers agreed to hold a referendum on the nation's dictatorship-era constitution. The constitutional referendum was demarked by two plebiscites: the first plebiscite (25 October 2020\footnote{The plebiscite was initially set for 26 April 2020. However, due to the COVID-19 pandemic, it was rescheduled for October of that year.}) asked whether a new constitution should be drafted; the second plebiscite (4 September 2022) was to vote on whether the people agreed with the text of the new constitution drawn up by the Constitutional Convention. These are popularly known in Chile as "entry plebiscite" (plebiscito de entrada) and "exit plebiscite" (plebiscito de salida).

In the entry plebiscite, the "Approve" side won by a large margin, with over 78\% agreeing to draft a new constitution. However, after two years of intense political campaigns from both sides, including heated social media discussions, the new text was rejected in the exit plebiscite with almost 62\% of the votes for "Reject".

\subsection{Data collection}
\label{chile_referendum}
Twitter is one of the most popular social media platforms in Chile\footnote{\url{https://www.statista.com/topics/6985/social-media-usage-in-chile/\#dossierKeyfigures}} for news consumption and where millions of Chileans discuss trending topics every day. Thus, we use Twitter as our source for topics and users' information. All data is collected using the official API.  

We start from a database of 384 news outlets with an active Twitter presence and targeting a Chilean audience \cite{epjds.s13688-019-0194-8}. 
Then, we collect tweets and profiles from these news outlets' followers. By focusing our analysis on people that consume their news from this media system, we target informed users that probably have a formed opinion on the discussed topics. We expect these users to leave traces of their stand, even if not explicitly shared online. However, we limit our analysis to accounts that follow at most ten different news outlets simultaneously to exclude potential automatic accounts.

Also, we use the \emph{location} field in the users' profiles to restrict the network to followers self-geolocated in one city, i.e., the capital of Chile, Santiago. This way, we try to minimize possible bias introduced by geographic and social factors. 

To further eliminate potential noise in the opinions, we restrict hyperactive accounts that, e.g., might be managed by automatic processes (i.e., bots) or work as part of an information campaign. Since these accounts usually do not represent real individuals, they will not convey a genuine personal instance within a controversial discussion. So, we introduce an additional filter based on the average daily number of tweets an account posts. Here we empirically chose at most three tweets per day on average as a reasonable activity level for a regular personal account. 

Note that users who don't participate in the referendum topic (the target topic in our case study) may be active in other discussions and regularly tweet about those other topics.

After applying the filters above, our first dataset ($Entry\_DS$) comprises 34,412 users with 915,672 associated tweets~(between Jan $1^{st}$ and October $24^{th}$, 2020) containing 189,115 hashtags. This dataset tries to capture the popular discussions during the political campaigns for the "entry plebiscite".


For our second dataset ($Exit\_DS$), we start with the same set of news outlets' followers. Then, for each account, we collected all tweets between Jan $1^{st}$ and September $3^{rd}$, 2022. After applying the same filters, our final $Exit\_DS$ comprises 39,239 users with 2,161,806  associated tweets containing 69,892 hashtags. Equivalent to the first dataset, $Exit\_DS$ tries to capture the popular discussions during the political campaigns for the "exit plebiscite".

\section{Methodology}
This section introduces the key elements of our methodology and describes the different components of the proposed model. 
First, we describe the pre-processing steps and hashtag classification. Then, we present the model's general architecture and discuss the integration of user-hashtag interactions. Following, we take on other types of information from social media interactions, including hashtag embeddings and the inferred relationship between users. Finally, we introduce optimizing the objective function of our model.

Our approach to predicting users' stance is through their affinity to hashtags that may represent this stance. For this, we represent the User-Hashtag relationship as a bipartite graph $\mathcal{G}_b$. The graph consists of two classes of nodes $\mathcal{V}_U$ and $\mathcal{V}_{HT}$, which represent the users and hashtags, respectively. A set of weighted edges $\mathcal{E}$ is defined to represent the interactions between users and hashtags. Then, each edge only connects nodes from different classes. We define the weight of an edge as $e_{i,j} = \frac{T_{i,j}}{\sum_j T_{i,j}}$, where $T_{i,j}$ is the number of times user $i$ used the hashtag $j$.

\subsection{Data processing}
\label{cleaning}
From the collected datasets, we normalize and standardize the hashtags encoding into UTF-8 and get 185,965 and 68,331 unique hashtags for $Entry\_DS$ and $Exit\_DS$, respectively. In addition to their selection by the users, hashtags' semantic information plays an important role. Therefore, we apply the following steps to process the content of the tweets:

\begin{itemize}
\item {\verb|standardization|}: encode all texts into UTF-8, replace the accented characters with regular ones (i.e., \'a $\rightarrow$ a), and lowercase the texts;
\item {\verb|removal|}: remove all URLs, emojis, punctuation, stopwords, as well as personal information like E-mails;
\item {\verb|lemmatization and stemming|}: tokens are lemmatized and stemmed into declined forms;
\item {\verb|word-embedding|}: we use the cleaned content to train a word embedding model\footnote{We use FastText with CBOW \cite{bojanowski2017enriching}}. From this, we get the representation of the hashtags in a latent space.
\end{itemize}

\subsection{Hashtag-based Stance Classification}
\label{ht_stance_annotate}
Previous works have used the attached hashtags to infer the stance of a tweet~\cite{di2021content}. However, inspecting all hashtags manually for multiple topics can still be expensive and time-consuming (e.g., we collected over 250K hashtags). Moreover, new hashtags may be introduced in the public discussion day-to-day.


For evaluation purposes, we manually annotate only a small subset of hashtags, especially those related to the Chilean referendums. These hashtags have been annotated by one of the authors, who is a native Spanish speaker. We inspect the 400 most used hashtags from each dataset and assign them to several topics. Furthermore, we split the hashtags related to the Chilean referendum into three groups: "\emph{POS}" indicating a favorable stance, "\emph{NEG}" indicating a rejecting stance, and "\emph{NEUTRAL}" indicating interest or engagement but with a neutral stance. We use these annotations in the validation step to measure the performance of our model (see Section ~\ref{experiments}). 
For the list of referendum-related hashtags, see Appendix \ref{sec:annotated_hashtags}.

Finally, we assign each user to one of the defined stances on a topic (e.g., \emph{POS}, \emph{NEG}, \emph{NEUTRAL}). To decide the stance of user $u_i$ on the referendum topic, we use the predicted affinities between this user and the annotated referendum-related hashtags. We select as stance the class with the greatest average affinity (see Equation \ref{stance_class}). 

\begin{equation}\label{stance_class}
\begin{aligned}
    stance(u_i) &= \underset{c \in C}{\arg \max} \frac{1}{\left | c \right |}\sum_{j \in c} \hat{y}(u_i,ht_j)
\end{aligned}
\end{equation}
where $C$ is in (\emph{POS}, \emph{NEG}, \emph{NEUTRAL}).
\subsection{Overall structure of our model}

\begin{figure}[t]
  \centering
  \includegraphics[width=\linewidth]{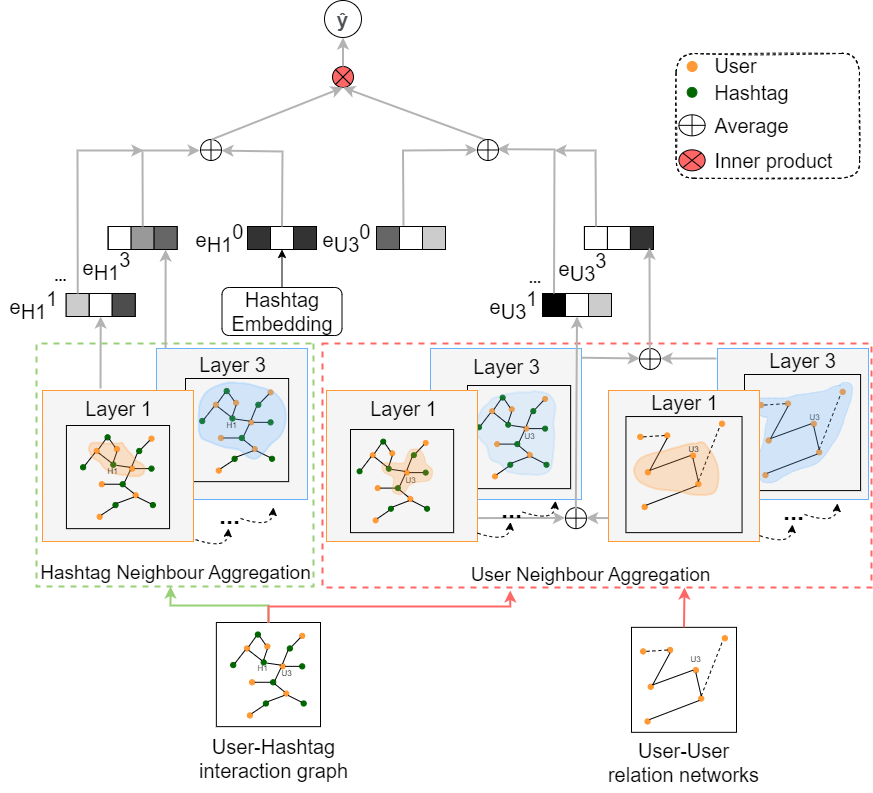}
  \caption{Overall architecture of the WLGCN model.}
  \Description{Schematic representation of the components of the proposed model and their interactions.}
  \label{fig:overall_architecture}
\end{figure}

Figure \ref{fig:overall_architecture} gives an overall view of our model. This represents an extension of the LightGCN that introduces weights to the relation graphs and various additional characteristic features of our social network. We will refer to our model as WLGCN. The model's inputs include a user-hashtag interaction graph, hashtag embeddings, and the inferred relationship between users. The output represents the users' predicted affinity to the hashtags in the dataset. Through a series of graph convolutional layers, the model jointly updates the representations of users and hashtags by aggregating the neighbors' features. After $K$ layers, the affinity score is calculated as the inner product of the users' and hashtags' embedded representation.

\subsection{Graph Convolutional Network}
\label{section_gcn}
The basic idea of Graph Convolutional Networks~(GCN) is to learn representations of nodes by aggregating the neighbors' embeddings as the new presentation of the target node. The layer-k embeddings of the target node can be represented as:
\begin{equation}\label{eq:neigh_emb}
\mathbf{h}_n^{k}=\mathrm{AGG}\left(\mathbf{h}_n^{(k-1)},\left\{\mathbf{h}_i^{(k-1)}: i \in \mathcal{N}_n\right\}\right), \quad \mathbf{h}_n^0=\mathbf{e}_n
\end{equation}
where $\mathbf{e}_n$ represents the initial embeddings of a node $n$, $\mathcal{N}_n$ represents neighbors of this node, and $AGG$ is a function used to aggregate the features of the neighbors. The other standard operations in a GCN layer (i.e., non-linear activation and feature transformation) have been shown to contribute little to the recommendation performance~\cite{he2020lightgcn}. Therefore, we also skip these two operations and use the simple average aggregator instead. To illustrate, consider our interaction graph $\mathcal{G}_b$ with $N$ users and $M$ hashtags~(HT)\footnote{We apply the same strategy to the other inferred graphs.}, the propagation rule in layer $k$ can be defined as:
\begin{equation}
    H^k=\left(D^{-\frac{1}{2}}AD^{-\frac{1}{2}}\right) H^{k-1}, \quad H^0=\mathbf{E}^0
\end{equation}
where $H^k \in \mathbb{R}^{\left(N+M\right) \times d}$ is the User-HT graph embedding matrix after the $k^{th}$ propagation step; $E^0$ is the initial $d$ dimensional embedding of users and HTs;  $D$ is a diagonal matrix, where $D_{i,i}$ equals to $\sum_j A_{i,j}$, $A$ stands for the User-HT graph adjacency matrix and is defined as:
\begin{equation}
    \mathbf{A}=\left(\begin{array}{cc}
\mathbf{0} & \mathbf{R} \\
\mathbf{R}^T & \mathbf{0}
\end{array}\right)
\end{equation}
being $R \in \mathbb{R}^{\left(N+M\right) \times d}$ the User-HT interaction matrix, where $R_{i,j} = e_{i,j}$ (i.e., the weight of the edge connecting user $i$ and hashtag $j$). 
After propagation, for the node $n$, which represents the user or the hashtag, we employ the weighted average to combine the embeddings learned through layers 1 to $K$, and the combination can be formulated as:
\begin{equation}
\label{pred_loss_func}
\begin{aligned}
E_{n} &=\frac{1}{K+1} \sum_{k=1}^K H_{n}^k
\end{aligned}
\end{equation}
Finally, we calculate the affinity by applying the inner product operations to the user and hashtag embeddings:
\begin{equation}
\hat{y}(u, ht)=e_{u} e_{ht}^{\top}
\end{equation}

\subsection{Inferred information}
\label{sec:multi-aspect_info}
The WLGCN model, described in Section~\ref{section_gcn}, focuses only on the user-hashtag interaction graph to jointly learn their representations, which is our main target. Here we introduce three additional types of data characteristic of our social network that can help in the above learning process.

First, we add hashtag embeddings to capture their semantics. The aim is to complement the hashtags usage patterns at the user level, represented by the vanilla WLGCN, with the contextual information provided by the tweets' content. For our experiments, we trained a FastText model \cite{bojanowski2017enriching} with the pre-processed corpus introduced in Section~\ref{sec:dataset}. Then, we use the representation of the hashtags ($E_{HT}$) to initialize the hashtag embedding layers of our model.

The second type of information is user-user network interactions. The user-user graph is an instance of a {\em Social Graph} with heterogeneous connections (henceforth $\mathcal{G}_{Soc}$). In $\mathcal{G}_{Soc}$, we include as links the mutual friend/follow relationship as well as mentions of and replies to other users in our network. 

The last type of information is the user-user simulated path (PathSim~\cite{sun2011pathsim}). The graph $\mathcal{G}_{Soc}$ mentioned above represents direct interactions observable from our Twitter dataset. However, in practice, these interactions are very sparse in a network like ours. So, we assume that they would offer a limited contribution to the embedding learning process. To address this issue, we extend these observed relations with inferred pseudo-relations based on \emph{meta-paths}. A \emph{meta-paths} captures a sequence of relations connecting two users that may contain multiple steps. For example, users $u_i$ and $u_j$ are connected through a path "user-retweet-hashtag-tweet-user"~(U-RT-HT-T-U) if $u_i$ retweeted a tweet containing a hashtag that also appeared in a tweet of $u_j$. Given the \emph{meta-paths}~($\mathcal{P}$ = U-RT-HT-T-U), the similarity between $u_i$ and $u_j$ is defined as:
\begin{equation}
s(i, j)=\frac{2 \times\left|\left\{p_{i \rightsquigarrow j}: p_{i \rightsquigarrow j} \in \mathcal{P}\right\}\right|}{\left|\left\{p_{i \rightsquigarrow i}: p_{i \rightsquigarrow i} \in \mathcal{P}\right\}\right|+\left|\left\{p_{j \rightsquigarrow j}: p_{j \rightsquigarrow j} \in \mathcal{P}\right\}\right|}
\end{equation}
where $p_{i \rightsquigarrow j}$ represent the path instance between $u_i$ and $u_j$ that follows the \emph{meta-paths} $\mathcal{P}$. In our previous example, the RT (retweet) relation can be replaced with other relations, such as reply. These path instances define an additional, denser user-user graph ($\mathcal{G}_{PathSim}$).
Since both graphs, $\mathcal{G}_{Soc}$ and $\mathcal{P}$ contained additional user information, We assume these graphs could be helpful in updating the embeddings of users. Inside each graph, we also applied Equation~\ref{pred_loss_func} with K layers to extract the potentially useful information. 

\subsection{Optimization}
With Equation~\ref{pred_loss_func}, our idea is to keep nodes connected with an edge close to each other in the latent space while pushing nodes without a shared edge farther apart. So, we adopt the Bayesian Personalized Ranking ~(BPR) loss~\cite{rendle2012bpr} as objectives for training our model:
\begin{equation}\label{bpr_loss}
    Loss=-\sum_{u=1}^N \sum_{i \in \mathcal{N}_u} \sum_{j \notin \mathcal{N}_u} \ln \sigma\left(\hat{y}(u, i)-\hat{y}(u, j)\right)+\lambda\left\|\mathbf{E}^0\right\|^2
\end{equation}
where $\sigma$ is the sigmoid function, $\lambda$ is the regularization parameter to avoid overfitting, and $i$, $j$ represent the hashtags that are used or not used by the user $u$. We adopt the Adam algorithm~\cite{kingma2014adam} for model optimization. We sample a tuple of ($u$, $i$, $j$) for each mini-batch and update the embeddings.


\section{Experiments and Results}
\label{experiments}

In this section, we first describe our experimental setup and evaluation approach. Then 
we contrast the performance of our model against several baselines. Finally, we analyze the impact of various levels of annotation effort. Across all our experiments, we use the Twitter datasets described in Section~\ref{sec:dataset} for our analyses.

\subsection{Experimental setup}
\subsubsection{Model Initialization}
Before training, at the first embedding layer, we use the Xavier uniform~\cite{glorot2010understanding} to initialize the embeddings of users. As for hashtags, the previously trained word embeddings are used for initialization. For comparison, we also try the hashtag representations with the Xavier uniform initializer in our experiments. For the number of convolutional layers K in our GCN, similar to previous works~(\cite{mei2021light}, \cite{he2020lightgcn}), we use three layers to extract and aggregate information from neighbor nodes. To prevent overfitting, early stopping is performed, i.e. the training will stop if recall@20 on the validation data does not increase for 50 successive epochs.

\subsubsection{Baselines}
We use two state-of-the-art methods as baselines to evaluate the performance of our proposal. In addition, we also use a Null-model to test whether the observed User-HT relations contain non-trivial information that helps in the identification of users' stances. Below we summarize the included baselines:

\begin{itemize}
\item {\verb|Null-Model|}~\cite{newman2004finding}: We create a randomized User-HT interaction matrix. We randomly sample $N$ interactions with replacement from a uniform distribution. $N$ is taken from the number of interactions (i.e., hashtag mentions) observed in the corresponding Twitter dataset.
\item {\verb|MF|}~\cite{rendle2020neural}: The Matrix Factorization method (MF) decomposes the User-HT interaction matrix into the product of two lower dimensionality matrices $U$ and $HT$. We use the MF implementation from ~\cite{rendle2020neural}.
\item {\verb|LightGCN|}~\cite{he2020lightgcn}: 
This GCN-based method simplifies the standard design of GCN to make it more concise and appropriate for collaborative filtering and recommendation tasks. It jointly learns user and item embeddings through a user-item interaction graph. Unlike our proposed WLGCM, LightGCN uses binary user-item interactions, while ours uses weights.
\end{itemize}

These baselines have different characteristics and cover different approaches to collaborative filtering. We compare these baselines against multiple variants of the proposed model to assess the contribution of its different aspects.

\subsection{Evaluation protocol}
\label{eval_protocol}
In the previous sections, we introduced the WLGCN model and the datasets used in this study. Here, we focus on the evaluation of the model's effectiveness in extracting useful information from the data. To this end, we test two key aspects: 1) the model's performance in predicting users' affinities toward each hashtag, thus, reflecting their preferences within a topic, and 2) the model's accuracy in predicting each user's overall stance on a topic, even in the absence of explicit knowledge about users' opinions for this particular topic.
These two aspects above translate into an investigation of the prediction performance of our model at two levels: edge and user level. To test the second aspect (user level prediction), we consider a specific topic: the Chilean constitutional referendum processes (2020 and 2022). For this, we rely on tweets that include referendum-related hashtags.

For our purposes, affinities are expressed as continuous values, with higher values indicating a stronger user preference. To assess the first point above (edge level performance), we use Recall@K and NDCG@K for user-hashtag interactions based on the top K recommendations with the highest affinities (K=20 in our experiments). 
We randomly select a fraction (5\% in our experiments) of the users that have interactions (i.e., edges in  $\mathcal{G}_b$) with the referendum-related hashtags and remove all these interactions. These removed edges are hidden from the model during training. They represent our ground truth and will be used later for testing. 
The remaining data (not used for testing) is utilized to evaluate the edge level performance via 5-fold cross-validation. We split this data into training and validation sets in a proportion of 80\% - 20\% and report the average recall and NDCG values over the five runs. 

For evaluation of the user level predictions, we use the hashtags interactions removed before (5\% of the users participating in the referendum topic). Starting from the trained models from the previous cross-validation analyses, we predicted affinities for the removed edges and compared them to the observed hidden interactions. Here, the stances of users from both the ground truth and the predicted affinities are computed using Equation~\ref{stance_class}.

For the user level, besides the user's stance prediction accuracy, we measure the root mean square error (RMSE) \cite{wang2019silent}.  
Since we are representing the stances in a continuous space but are evaluating the accuracy with discrete values or classes (i.e., negative, neutral, positive), RMSE helps us assess how far or close our predictions are to the ground truth classes before applying the transformation in Equation~\ref{stance_class}.
Note that a smaller RMSE or a higher accuracy value indicates a better inference performance in the experiments. 

%


\subsection{Comparison with Baselines}
\begin{table}
\caption{Edge- and User-level Performances. LightGCN uses binary values in the interaction graph, while WLGCN uses weighted values.}
\label{tab:results}
\small
\centering
\begin{tabular}{@{}l|rr|rr}
\toprule
\multicolumn{1}{c}{$Entry\_DS$}  & \multicolumn{2}{c}{Edge}                                & \multicolumn{2}{c}{User}  \\ \cmidrule(l){2-5} 
\multicolumn{1}{c}{}                                   & \multicolumn{1}{c}{Recall} & \multicolumn{1}{c}{NDCG} & \multicolumn{1}{c}{Acc.} & \multicolumn{1}{c}{RMSE}\\ \midrule
Null Model &  1e-5 & 1e-5 & 0.253 & 0.864 \\
MF  & 0.267 & 0.201 &  0.438 & 0.75 \\
LightGCN & 0.180 & 0.127 & 0.801 & 0.446 \\
WLGCN & 0.269 & 0.191 & 0.814 & 0.431 \\
WLGCN + ($E_{HT}$)  & 0.274 & 0.198 & \textbf{0.818} & \textbf{0.427}\\
WLGCN + ($\mathcal{G}_{PathSim}$)   & 0.276 & 0.199 & 0.805 & 0.442\\
WLGCN + ($E_{HT}$, $\mathcal{G}_{PathSim}$)  & 0.274 & 0.201 & 0.811 & 0.434\\
WLGCN + ($\mathcal{G}_{Soc}$) & \textbf{0.279} & \textbf{0.202} & 0.806 & 0.44\\
WLGCN + ($E_{HT}$, $\mathcal{G}_{Soc}$)  & 0.275 & 0.201 & 0.811 & 0.434\\
WLGCN + ($\mathcal{G}_{PathSim}$, $\mathcal{G}_{Soc}$)       &  0.274 & 0.199 & 0.804 & 0.442\\
WLGCN + ($E_{HT}$, $\mathcal{G}_{PathSim}$, $\mathcal{G}_{Soc}$) & 0.263 & 0.199 & 0.806 & 0.44\\
\midrule
\multicolumn{1}{c}{$Exit\_DS$} & \multicolumn{2}{c}{Edge}                                & \multicolumn{2}{c}{User} \\ \cmidrule(l){2-5} 
\multicolumn{1}{c}{}                                   & \multicolumn{1}{c}{Recall} & \multicolumn{1}{c}{NDCG} & \multicolumn{1}{c}{Acc.} & \multicolumn{1}{c}{RMSE}\\ \midrule
Null Model & 8e-4 & 3e-4 &  0.524 & 0.69\\
MF  & 0.26 & 0.125 & 0.615 & 0.62\\
LightGCN & 0.207 & 0.125 & 0.66 & 0.583\\
WLGCN & 0.232 & 0.126 & 0.673 & 0.572\\
WLGCN + ($E_{HT}$) & 0.26 & 0.128 & 0.692 & 0.555\\
WLGCN + ($\mathcal{G}_{PathSim}$)  & \textbf{0.266} & \textbf{0.132} & 0.691 & 0.556\\
WLGCN + ($E_{HT}$, $\mathcal{G}_{PathSim}$) & 0.262 & 0.129 & \textbf{0.694} & \textbf{0.553}\\
WLGCN + ($\mathcal{G}_{Soc}$) & 0.265 & \textbf{0.132} & 0.692 & 0.555\\

WLGCN + ($E_{HT}$, $\mathcal{G}_{Soc}$) & 0.263 & 0.13 & 0.692 & 0.555 \\
WLGCN + ($\mathcal{G}_{PathSim}$, $\mathcal{G}_{Soc}$)       & 0.264 & 0.131 & 0.682 & 0.564\\
WLGCN + ($E_{HT}$, $\mathcal{G}_{PathSim}$, $\mathcal{G}_{Soc}$) & 0.263 & 0.13 & 0.690 & 0.557\\ 
\bottomrule
\end{tabular}
\end{table}


In this study, we evaluate the performance of four models, including the three baselines (null model, MF, LightGCN) and the proposed WLGCN, using four evaluation metrics: recall@20, NDCG@20, accuracy, and RMSE. The results are summarized in Table~\ref{tab:results}. 

From our experiments, we first notice that WLGCN significantly outperforms the null models at both levels. This indicates that our model is able to extract meaningful features from the input interactions, such as community structures. 
It is also noteworthy that MF, at the edge level, proves to be highly competitive, surpassing LightGCN and even the vanilla version of WLGCN in recall~($Exit\_DS$) or NDCG~($Entry\_DS$). Nevertheless, LightGCN provides more accurate user attitudes predictions than MF and the null model. However, extended versions of WLGCN outperform LightGCN and MF in all four metrics at both levels. 

Overall, the results in Table~\ref{tab:results} show that, although MF has a competitive performance at the edge level, the models based on WLGCN are more reliable across both levels and with different datasets.
Note that, in modeling users' attitudes, both tasks are essential. While the edge level prediction identifies the most interesting topic(s) for the user, the user level requires a more global representation of their preferences in all subjects. For example, if a user is not very interested in politics, (s)he will probably not have the referendum-related hashtags at the top of her affinities. Nevertheless, based on other choices and connections, our model needs to roughly rank these hashtags so that, on average, we get the user's leaning within the topic.
The more stable performance of WLGCN in all four metrics indicates that it is the most effective model for addressing hashtag-based stance prediction.

\subsection{Effect of inferred information}
\label{sec:effect_inferred_info}
The proposed WLGCN shows that a weighted User-HT interaction matrix improves the outcomes of our tasks compared to previous approaches. However, Table~\ref{tab:results} suggests that integrating additional information sources into the WLGCN model further enhanced its performance. Performances on the first dataset showed that the combination of WLGCN and $\mathcal{G}_{Soc}$ produced the best results at the edge level. In contrast, the combination of WLGCN and $E_{HT}$ made for the highest accuracy at the user level. These resulted in improvements of 4.5\% in recall and 2.1\% in accuracy, respectively. Note that the social graph ($\mathcal{G}_{Soc}$) contributes extra knowledge on the users. On the other hand, the semantic information from hashtags ($E_{HT}$) provides additional associations between hashtags. Interestingly, the combination of these two seemingly complementary dimensions does not produce a better model but rather an intermediate result between the two features alone.

For the second referendum, both the combination of WLGCN + $\mathcal{G}_{Soc}$ and the variation of WLGCN + $\mathcal{G}_{PathSim}$ produced similarly optimal results at the edge level, with a 2.3\% improvement in recall over the baselines. However, the combination of WLGCN + $E_{HT}$ again performed best at the user level with a 5.2\% improvement in accuracy. Although, the alternatives of WLGCN and any of the social graphs ($\mathcal{G}_{Soc}$ or $\mathcal{G}_{PathSim}$) were still more competitive than for the first referendum.

An interesting outcome is the comparison between the two social graphs as complements to the WLGCN. Even though the $\mathcal{G}_{PathSim}$ is significantly less sparse than the $\mathcal{G}_{Soc}$ (initial motivation to include it in our model), their performances remain very close in both datasets. This suggests that the more direct user-user relations (e.g., followers, mentions, replies), albeit less frequent, are more relevant and able to capture as much information as the simulated paths.

In general, our results with the variations of WLGCN indicate that aggregating multiple types of information is not straightforward. This is demonstrated by the combination of the best-performing feature with other relationships, which resulted in a decrease or similar performance in most cases. The fitting mechanism for combining the intermediate representation from these features is a challenging task that we will explore further in future research. So far, we have tried various methods, such as attention. However, averaging them produced consistently superior results and thus was the preferred method for the reported analyses.

Nevertheless, overall, the results of our experiments show that the combination of WLGCN and other features produces improved predictions of user opinions and affinity scores with varying degrees of effectiveness depending on the information source and evaluation metric. These results provide a foundation for future research in this area.

\subsection{Annotation effort analysis}
\begin{figure}[t!]
  \centering
  \includegraphics[width=\linewidth]{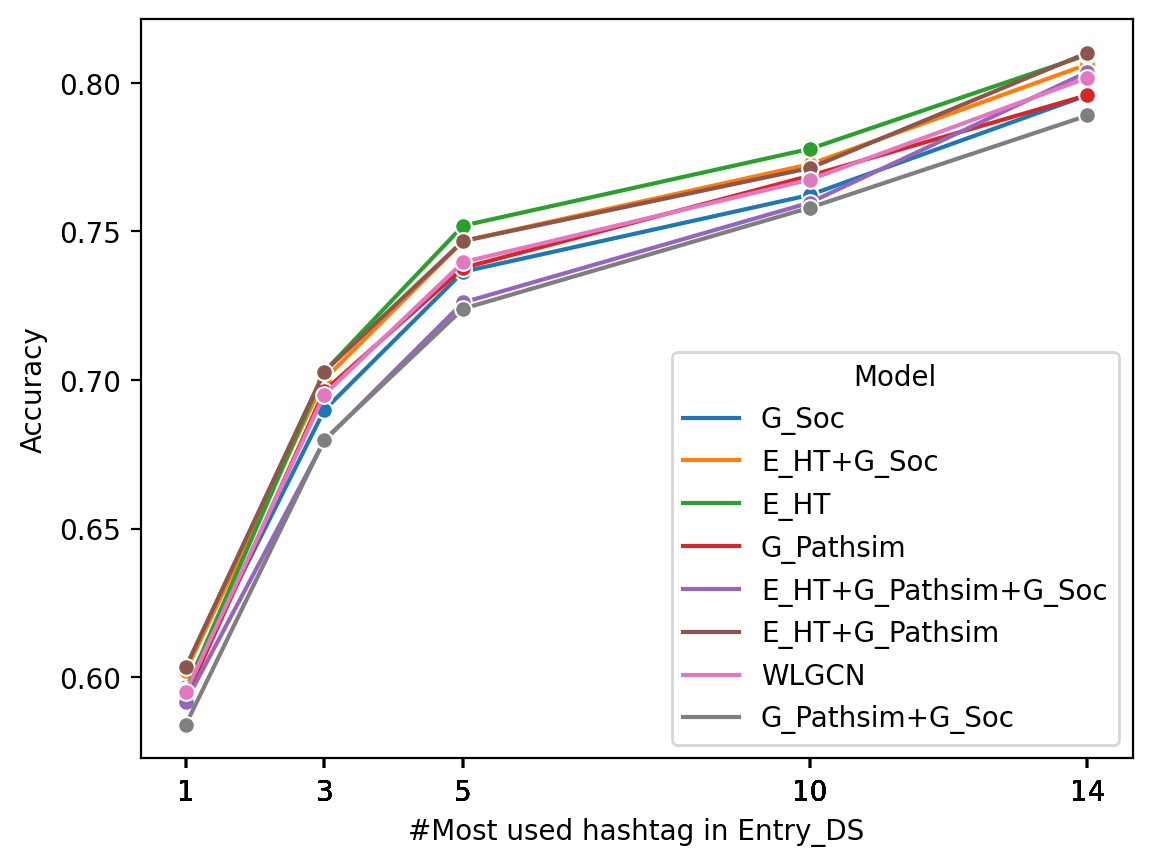}
  \caption{User-level accuracy using different hashtag annotations in $Entry\_DS$}
  \label{fig:acc_entry}
\end{figure}

\begin{figure}[t!]
  \centering
  \includegraphics[width=\linewidth]{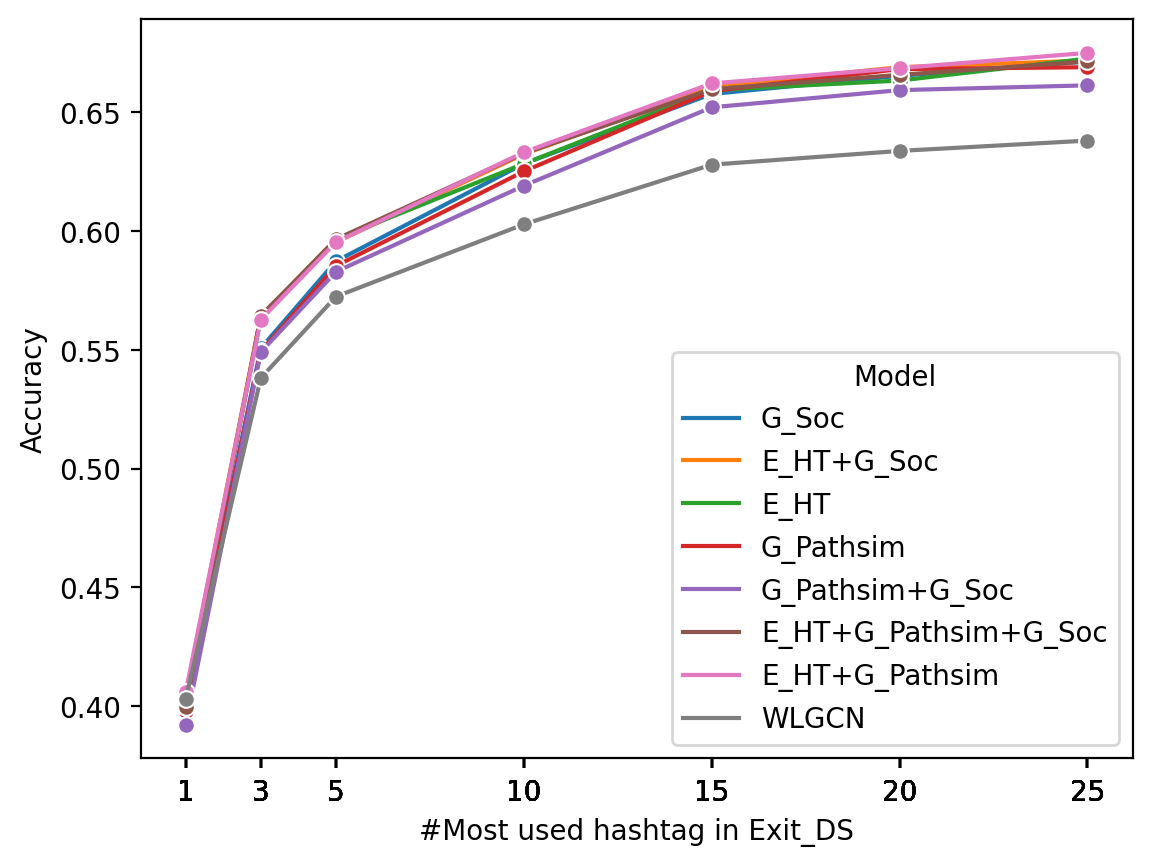}
  \caption{User-level accuracy using different hashtag annotations in $Exit\_DS$}
  \label{fig:acc_exit}
\end{figure}

Another advantage of our approach compared to previous works is the minimal annotation required (e.g., \cite{wang2019silent}). As presented before, only a set of hashtags related to the topic of interest must be identified. Still, the model is able to profit from other discussions and interactions potentially outside this topic.
To further investigate the impact of an increased expert effort, we experiment with user level stance prediction and a growing number of annotated hashtags. The results are shown in Figure~\ref{fig:acc_entry} and Figure~\ref{fig:acc_exit} for the $Entry\_DS$ and $Exit\_DS$, respectively.

In Table~\ref{tab:entry_annotated_referendum_hts} (Appendix~\ref{sec:annotated_hashtags}), we include the referendum-related hashtags (annotated by the authors). For the results discussed in the previous section, we used all of them to evaluate the models on a per-user basis (i.e., estimated affinity from each user to all hashtags). The focus of this section is to examine the variability of accuracy in the proposed model at the user level when different numbers of hashtags are annotated. As shown in Figures ~\ref{fig:acc_entry} and ~\ref{fig:acc_exit}, the $x$-axis represents a prediction of the users' stances when including only $x$ annotated referendum-related hashtags for each stance class ($1 \leq x \leq \min(|POS|,|NEG|)$).
Note that we have ($|POS|=14, |NEG|=21, |NEU|=5$) in $Entry\_DS$ and ($|POS|=26, |NEG|=25, |NEU|=4$) in $Exit\_DS$. So, we don't consider the neutral ones because only a few were found.

In each case, we select the top-$x$ most used hashtags in each class. These should represent the easiest ones to identify by the experts and thus require the least effort. For example, $x=3$ means that from each class in {\emph{POS}, \emph{NEG}}, we chose for the calculation in Equation~\ref{stance_class} only the top-3 most used hashtags related to the referendums.


Both figures show similar behavior. As expected, a higher number of annotations leads to higher accuracy. However, the increase in accuracy slows down after five hashtags and tends to become asymptotic as the number of annotations increases, especially in $Exit\_DS$. This tendency strengthens the practical implications in the applicability of our model as it could further simplify our approach. For example, we might only know some of the related hashtags for a new topic, or they could evolve over time. An expert could only need to annotate a small sample of the most used hashtags related to that topic. As a result, the performance should remain stable without heavy annotation work.

\section{Conclusions}
In conclusion, this study aimed to predict the attitude of social media users on selected topics by combining content and social interaction. To tackle this problem, we proposed a collaborative filtering model based on Graph Convolutional Networks that estimates the users' affinity to hashtags that represent stances on a discussion. Furthermore, considering the sparsity of user-hashtag / user-topic interactions, we explored the impact of different relationships between elements on the embedding update process and final predictions. The experiments were conducted using two large datasets collected from Twitter during two Chilean referendums and showed the effectiveness of our approach compared to state-of-the-art baselines. 

Our results show that supplementary knowledge from hashtags' semantic information ($E_{HT}$) or users' social interactions ($\mathcal{G}_{Soc}$ or $\mathcal{G}_{PathSim}$) positively impact the performance of our model. However, combining multiple of these extra features proved to be a challenging task.
For example, the most useful inferred information varies for different scenarios. We could not find a silver bullet that produced the best results in all tested cases. Thus, future work will explore ways to incorporate multiple relationships between users and other relevant information effectively.

 Also as future work, we are interested in identifying and tracking the shifts in users' opinions over time. Other lines of research extending this work include the generalizability of the prediction to the offline public and stance characterization for discussions with multiple poles.

\begin{acks}
This paper is part of a project that has received funding from the \grantsponsor{EU}{European Union}{https://doi.org/10.3030/101021866}'s Horizon 2020 research and innovation programme under grant agreement No. \grantnum[]{EU}{101021866} (CRiTERIA). 
\end{acks}

\bibliographystyle{ACM-Reference-Format}
\bibliography{sample-base}

\appendix
\newpage
\section{Annotated referendum-related hashtags}
\label{sec:annotated_hashtags}

\begin{table}[h]
  \caption{Annotated referendum-related hashtags}
  \label{tab:entry_annotated_referendum_hts}
  \begin{tabular}{c|p{6cm}}
\toprule
Stance  & $Entry\_DS$ \\
    \midrule
     \emph{POS} & apruebo, apruebo26abril, apruebocc, apruebochiledigno, aprueboconvencionconstitucional, aprueboganaenoctubre, apruebonuevaconstitucion, apruebosinmiedo, nuevaconstitucionparachile, yoapruebo, yoapruebocc, yoapruebolanuevaconstitucion, yoapruebonuevaconstitucion, yovotoapruebo \\
     \midrule
     \emph{NEG} & lacallerechaza, noalanuevaconstitucion, porchileyorechazo, rechazo, rechazocrece, rechazoganaenoctubre, rechazoganasivotamos, rechazoganasivotamostodos, rechazonuevaconstitucion, rechazoporchile, rechazosalvaachile, rechazosalvachile, rechazosinmiedo, rechazotutongo, rechazoynulo, retrazo, votarechazo, votorechazo, yorechazo, yorechazonuevaconstitucion, yovotorechazo \\
    \midrule
    \emph{NEUTRAL} & convencionconstitucional, convencionconstituyente, nuevaconstitucion, plebiscito2020, plebiscitochile \\
  \bottomrule
Stance  & $Exit\_DS$\\
\midrule
     \emph{POS} & aprobamosfelices, aprobareshumano, apruebaserahermoso, apruebaxchile, apruebazo, apruebo, apruebo4deseptiembre, aprueboconesperanza, apruebocrece, apruebodesalida, aprueboel4deseptiembre, apruebofeliz, apruebonuevaconstitucion, aprueboparaquenuncamasenchile, aprueboplebicitodesalida, apruebosincondiciones, apruebosinmentiras, apruebosinmiedo, apruebounchilemejor, aprueboxamor, chilevotaapruebo, laconvencionsedefiende, mivotonocambia, rechazoganael4deseptiembre, yoapruebo, yoapruebofeliz\\
          \midrule
     \emph{NEG} & circoconstituyente, convencionculia, rechazo, rechazoconesperanza, rechazoconfuerza, rechazocontodos, rechazocrece, rechazodesalida, rechazodesalida2022, rechazoel4deseptiembre, rechazoelmamarracho, rechazoelmamarrachocomunista, rechazoelplurimamarracho, rechazoganael4deseptiembre, rechazoladestrucciondechile, rechazopopular, rechazoporamorachile, rechazoporchile, rechazosalvaachile, rechazosalvachile, rechazotransversal, rechazoxamorachile, rechazoxchile, rechazoypunto, yorechazo\\
          \midrule
    \emph{NEUTRAL} & convencionconstitucional, convencionconstituyente, nuevaconstitucion, plebiscitodesalida\\
  \bottomrule
\end{tabular}
\end{table}

\end{document}